\documentclass[traditabstract]{aa}

\usepackage{graphicx}
\usepackage{txfonts}

\def\PKS1830{PKS\,1830$-$211}

\begin{document}

\title{Submillimeter polarization and variability of quasar \PKS1830}

\author{I. Marti-Vidal \and 
        S. Muller\inst{1}
}

\offprints{I. Marti-Vidal \\ \email{mivan@chalmers.se}}
\institute{Department of Space Earth and Environment, Chalmers University of Technology, 
           Onsala Space Observatory, SE-43992 Onsala, Sweden
}

\date{Accepted for publication in A\&A}
\titlerunning{Polarization in \PKS1830 at 2.3\,THz}
\authorrunning{Marti-Vidal et al. (2018)}

\abstract{Polarization from active galactic nuclei is interpreted as a signpost of the role of magnetic fields in the launch and collimation of their relativistic radio jets.
Here, we report the detection of a clear polarization signal from ALMA observations of the gravitationally lensed quasar \PKS1830\ at submillimeter wavelengths (Band 9, 650~GHz). Applying a differential-polarimetry technique to the two compact lensed images of the quasar, we estimate a fractional polarization of $\sim 5$\% for one lensed image, while the other appears nearly unpolarized, which implies that the polarization activity varies on a timescale of a few weeks. With additional ALMA Band 7 and 8 (between 300--500~GHz) concomitant data, we constrain a Faraday rotation of a few $10^5$~rad\,m$^{-2}$. We also observe flux-density variability of $\lesssim 10$\% within one hour in Band~9. This work illustrates that a differential analysis can extract high-accuracy information (flux-density ratio and polarimetry) free of calibration issues from resolved sources in the submillimeter domain.}


   \maketitle

\section{Introduction}

Understanding the origin of polarization from active galactic nuclei (AGN) and its correlations with the emission across the electromagnetic spectrum (from radio to $\gamma$-rays) is currently one of the greatest challenges in the observational and theoretical studies of AGN jets. Polarization is a sensitive probe of magnetic fields, which may play a critical role in the formation and propagation of the relativistic outflows from AGN. Changes in the electric-vector position angle (EVPA), which imply changes in the magneto-ionic conditions, either internal to the jets or in their immediate neighborhood, have been observed with timescales from days to months, and from optical \citep[e.g.,][]{kikuchi1988} to radio \citep[e.g.,][]{Marscher2008}. There is also evidence of EVPA correlations between optical and radio, especially at high frequencies \citep{Algaba2012}, and correlations between  EVPA at radio and $\gamma$-ray activity have also been reported \citep[e.g.,][]{Marscher2008}. 

Given the high magnetic fields and plasma densities close to the base of the AGN jets \citep[i.e., the place where the accreted matter is fed into the outflow by means of processes that are not yet fully understood; e.g.,][]{BlandfordZnajek}, the synchrotron emission from these regions is blocked by self-absorption at low radio frequencies \citep[e.g.,][]{Blandford1979}, which implies the need for observing at higher frequencies (e.g., the millimeter to submillimeter, mm/submm, domain), to probe the magnetic conditions and processes related to the onset of relativistic outflows. Hence, polarization observations are crucial at mm/submm wavelengths for studying AGN jet formation, collimation, and acceleration. 

\PKS1830 is a gravitationally lensed quasar at a redshift of $z=2.5$ (\citealt{lid99}). It shows two bright and compact images (one NE and the other SW) that are separated by $1\arcsec$ in the sky and are embedded in a patchy circular structure that is qualitatively similar to an Einstein ring seen at radio wavelengths \citep{jau91}. The time delay between the two images is $\sim 27$~days (\citealt{lov98,wik01,bar11}), with the NE image leading. Strong polarization (up to $\sim 20$\%) has been reported at radio centimeter \citep{sub90} and millimeter wavelengths \citep{Garrett1998}, and a high rotation measure ($RM$) is reported at mm/submm wavelengths \citep{IMV2015}. This source is also a strong $\gamma$-ray emitter, with several flaring events reported in recent years by the Fermi-LAT collaboration \citep[e.g.,][]{Abdo2015}. 

\PKS1830 has previously been observed with ALMA, primarily to study the molecular absorption arising from the foreground lensing galaxy \citep[e.g.,][]{mul14}. However, we have also used these observations to study the AGN variability and polarization as a byproduct. \cite{IMV2013} noted rapid and chromatic time variations in the flux-density ratio between the two lensed images. This is a very sensitive diagnostic of the AGN variability and is virtually free of calibration biases and instrumental errors. These variations were concomitant with a $\gamma$-ray flare, and could be explained using a simplified model of traveling plasmon across the jet current.

Furthermore, \cite{IMV2015} noted that it is possible to recover some source polarization information, even from standard dual-polarization observations (i.e., not designed for a full-polarization study), using the polarization-wise ratio of the intensity ratios between the two lensed images; a quantity that we called $R_{pol}$,

\begin{equation}
R_{pol} = \frac{1}{2}\left(\frac{R_{XX}}{R_{YY}}-1\right) ~~~~ \mathrm{with} ~~~~ R_{XX} = \frac{S^{XX}_{NE}}{S^{XX}_{SW}},
\end{equation}

\noindent where $S^{i}_{k}$ is the flux density of the $k$-th source image, measured from the $i$-th polarization channel (i.e., either the X or Y orthogonal polarizer of the ALMA receivers). $R_{pol}$ is related to the difference in polarization between the NE and SW images, which can depend on frequency and time (i.e., source intrinsic variations) and on parallactic angle (instrument-related variations).  

With a set of observations close in time (i.e., within a day, to limit the impact of intrinsic source variability) and at several frequencies, it is furthermore possible to measure the Faraday rotation, $RM$, from the changes of $R_{pol}$ in both frequency and parallactic angle, $\psi$. \cite{Baobab2016} also used dual-polarization ALMA submm data to measure $RM$ of a few $10^5$~rad\,m$^{-2}$ for Sgr\,A*. However, their method (i.e., standard Earth-rotation polarization synthesis) relied strongly on the stability of the antenna amplitude gains, whereas a differential approach (i.e., based on relative intensities) is independent of gain variability and hence more robust. In addition, \cite{hov18} recently reported the first ALMA full-polarization measurement of $RM$, with $RM$ also of a few $10^5$~rad\,m$^{-2}$ for 3C\,273 at 1~mm. For \PKS1830, \cite{IMV2015} derived even higher $RM \sim (1-2) \times 10^7$\,rad\,m$^{-2}$ (values not corrected for redshift) between sky frequencies of 250~GHz and 300~GHz. Such high $RM$s imply a high magnetic field in the sub-parsec region close to the base of the AGN jet.

Here, we report on ALMA dual-polarization observations of \PKS1830 taken in Band~9 (650\,GHz; 2.3\,THz corrected for redshift), spanning about one hour on the target and covering a wide range of parallactic angles ($\Delta \psi \sim 120^{\circ}$). Based on these observations, we use the flux-density ratios between lensed images to analyze the short-term (intra-hour) source variability, and we use the $\psi$ dependence on $R_{pol}$ to analyze the source polarization. In addition, we use complementary observations at Bands~7 and 8 (at frequencies between 300 and 500\,GHz) to constrain the Faraday rotation. In the next section, we describe our observations and analysis approach. In Sect. \ref{sec:res} we present and discuss our results. In Sect. \ref{sec:summ} we summarize our conclusions.

\section{Observations and data analysis} \label{sec:obs}

The observations were carried out with the Atacama Large Millimeter/submillimeter Array (ALMA) in dual-polarization mode between May and June 2015 in the frame of different projects aiming at detecting various molecular species in the $z=0.89$ galaxy \footnote{acting as both lens and absorber} located in front of \PKS1830\ (e.g., \citealt{mul14}). A summary of the datasets is given in Table~\ref{tab:summary-obs}.

The correlator was set up with spectral windows 1.875~GHz wide, and channel spacings of 0.977~MHz or 1.953~MHz. For this work, which focuses on the quasar continuum, the channels with molecular absorption lines were flagged and the remaining channels were used to build the continuum. The data were calibrated within the CASA\footnote{http://casa.nrao.edu/} package, following a standard procedure. The bandpass response of the antennas was calibrated from observations of the bright quasar J\,1924$-$292. The absolute flux calibration was scaled from Titan or Ceres, using the Butler-JPL-Horizons 2012 model. The flux-density measurements of J\,1924$-$292 were cross-checked with the values from the ALMA calibrator database (obtained from extrapolation of Band 3 and 7 measurements after determining the spectral index) and the consistency was on the order of 10\%, that is, within the ALMA expectations of 10--20\% for absolute flux accuracy in Bands 7--9. The gain solutions were then self-calibrated on the continuum of \PKS1830, integration-based (6~s) for the phases and scan-based ($\sim 3-5$~min) for the amplitudes. In the calibration, we solved for common gain solutions for the two orthogonal polarizers.

After calibration, the flux densities of the two lensed images of \PKS1830\ were extracted using the CASA-python task UVMULTIFIT (\citealt{IMV2014}) by fitting a model of two point sources to the interferometric visibilities. After checking for consistency of the fitting of the XX and YY Band~9 visibilities observed on 2015 May 19, a clear polarization signal was identified from the difference of the recovered flux density between the two polarizers. This signal was also evident in the CLEANed images of the quasar. The approximately one-hour observations covered a wide and continuous range of parallactic angles between 100 and 230~deg. For further investigations, in particular to test variations with parallactic angle, the data were binned on short time intervals of a few minutes. Even on these short time bins, the high sensitivity and good instantaneous uv-coverage of ALMA allow us to derive sensible visibility-fit solutions. Although the observations were originally not designed for a polarization study, the differential polarization method eventually allows us to retrieve some polarization properties of the quasar. The related results are discussed in Sec.\ref{sec:resB9}.

Unlike Band~9 data, observations at other frequencies do not offer a large parallactic angle coverage. Still, as for the Band~9 data, we have binned them in short intervals of few minutes on source and performed the same visibility-fitting analysis to check for the stability of the method and source intrinsic variations. Results are shown in Sec.\ref{sec:resB78}.

\begin{table*}[ht]
\caption{Summary of the ALMA observations of \PKS1830\ in May-June 2015.}
\label{tab:summary-obs}
\begin{center} \begin{tabular}{lcccccc}

\hline
Date & Time range                      & Parallactic     & ALMA & Frequencies $^a$  & Feed Angle$^b$  & Flux / Bandpass    \\
     &  (UTC)                          & Angle ($^\circ$) & band & (GHz ) & ($^\circ$) & calibrator  \\
\hline
2015 May 14 & 06:31:59.9 -- 06:56:17.5 & 260 -- 257 & B8   & 403.5, 401.5, 413.8, 415.5  & 0 & Titan / J1924$-$292 \\ 
2015 May 19 & 07:10:30.7 -- 08:17:17.0 & 230 -- 100 & B9   & 647.7, 649.9, 651.7, 653.6  & $-180$ & Titan /J1924$-$292 \\ 
2015 May 19 & 10:22:56.5 -- 10:46:44.4 & 102 -- 101 & B7   & 327.0, 337.1, 339.0         & $-53.6$ & Ceres / J1924$-$292 \\ 
2015 May 20 & 08:17:14.6 -- 08:38:54.3 & 100 --  99 & B8   & 428.5, 430.3, 440.6, 442.4  & 0 & Titan / J1924$-$292 \\ 
2015 Jun 06 & 04:23:14.4 -- 04:43:52.1 & 261 -- 261 & B8   & 480.2, 482.0, 492.2, 494.0  & 0 & Titan / J1924$-$292 \\ 

\hline
\end{tabular} 
\tablefoot{$a)$ Frequencies at the center of each 1.875~GHz wide spectral window. $b)$ Angle of the X and Y polarizers with respect to the alt-az axes of the antenna mount.}
\end{center} \end{table*}

\section{Results} \label{sec:res}

\subsection{Band 9 data} \label{sec:resB9}

\begin{figure} \centering
\includegraphics[width=9cm]{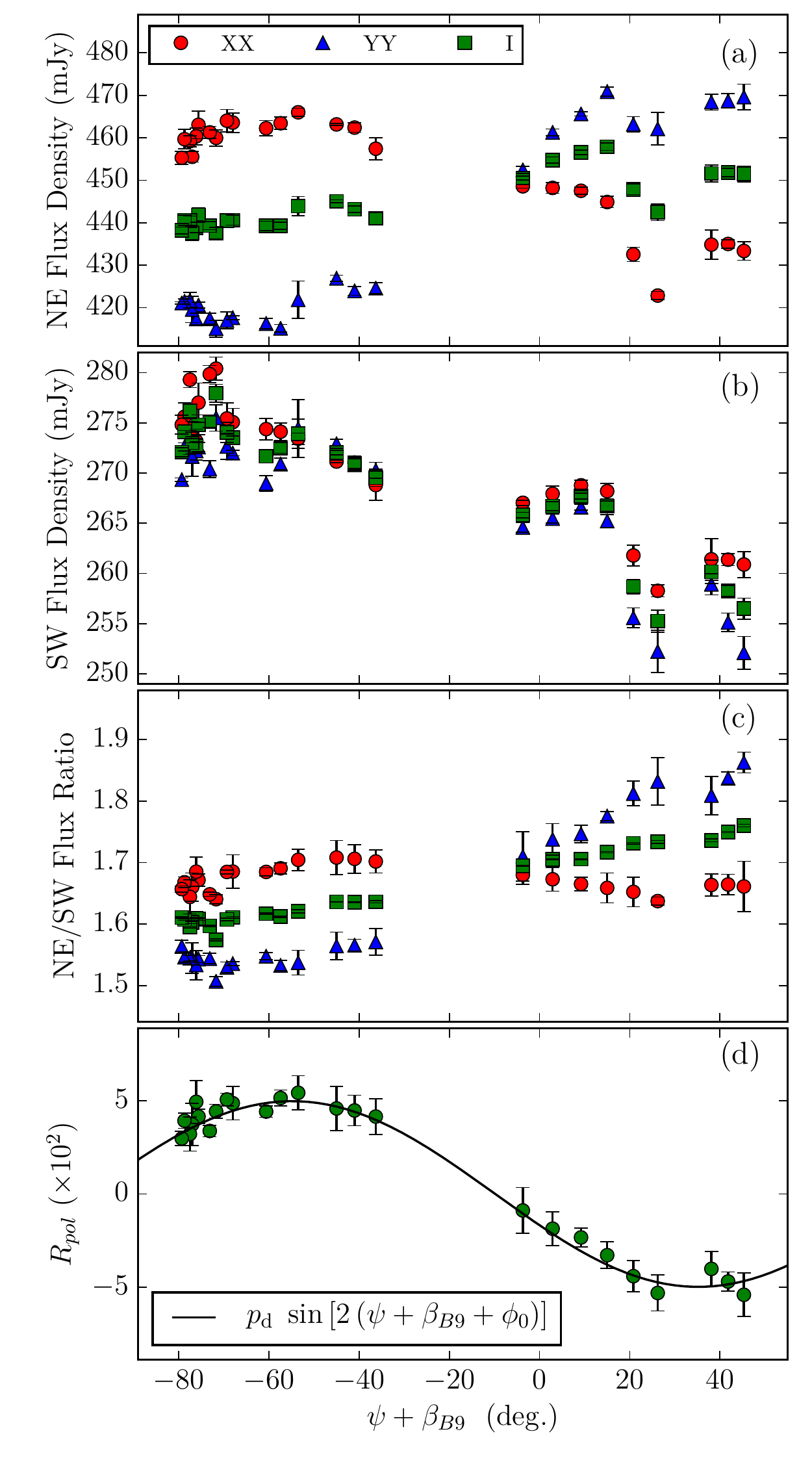}
\caption{From top to bottom: (a) flux density of the NE image (in XX and YY polarization products, as well as in Stokes I); (b) flux density of the SW image (same products as those of NE shown in (a)); (c) flux-density ratios between NE and SW; (d) polarization ratio, $R_{pol}$ (see text). All quantities are shown as a function of parallactic angle, $\psi$, rotated by the feed angle of the Band~9 receivers at ALMA, $\beta_{B9}$.}
\label{B9Fig} \end{figure}

The measured flux densities of the NE and SW images, their flux ratio, and $R_{pol}$ for the Band~9 dataset are shown in Fig.\ref{B9Fig} as  a function of the parallactic angle, $\psi$.

First, Fig.\ref{B9Fig}a shows a clear polarization signal from the NE image (i.e., different flux densities in XX and YY depending on the parallactic angle, $\psi$) that is not seen in the SW image (Fig.\ref{B9Fig}b). Since the angular separation between both images is only $1\arcsec$ in the sky (i.e., much smaller than the primary beam of $\sim 10 \arcsec$ of the ALMA antennas), it is difficult to explain this different behavior as instrumental and/or atmospheric effects.

On the other hand, the SW image seems to be nearly unpolarized (i.e., the flux densities in XX and YY are compatible with those in Stokes I across the range of $\psi$) and its flux density clearly increases in time compared to the relatively constant flux density of the NE image. Here, we note that while the absolute flux accuracy of the Band~9 data is expected to be on the order of 20\%, the relative flux accuracy of the data points in Fig.\ref{B9Fig}b and c depends only on the instrumental/atmospheric stability across the field of view.

\begin{figure*}
\centering
\includegraphics[width=6.4cm]{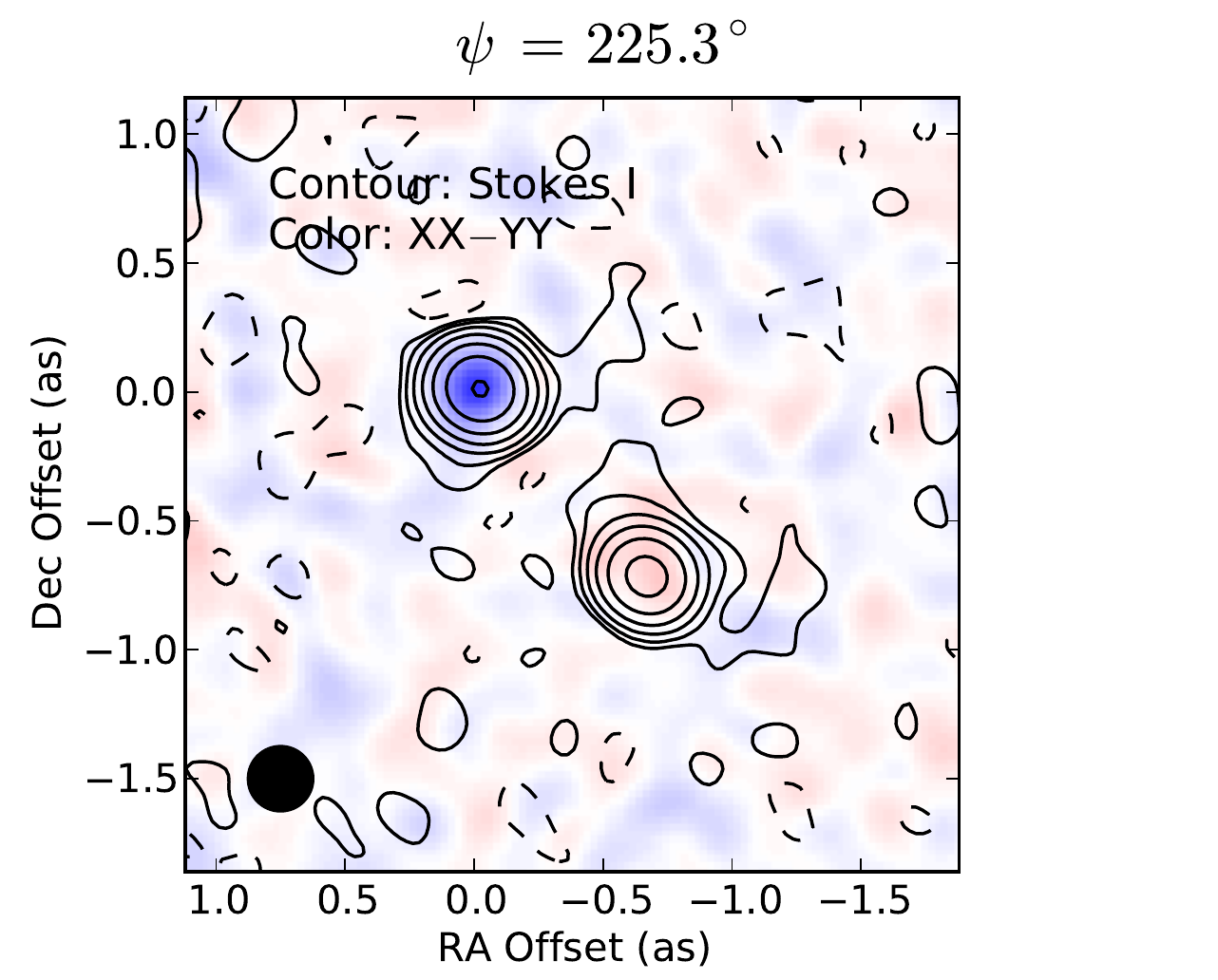} \hspace{-1.5cm}
\includegraphics[width=6.4cm]{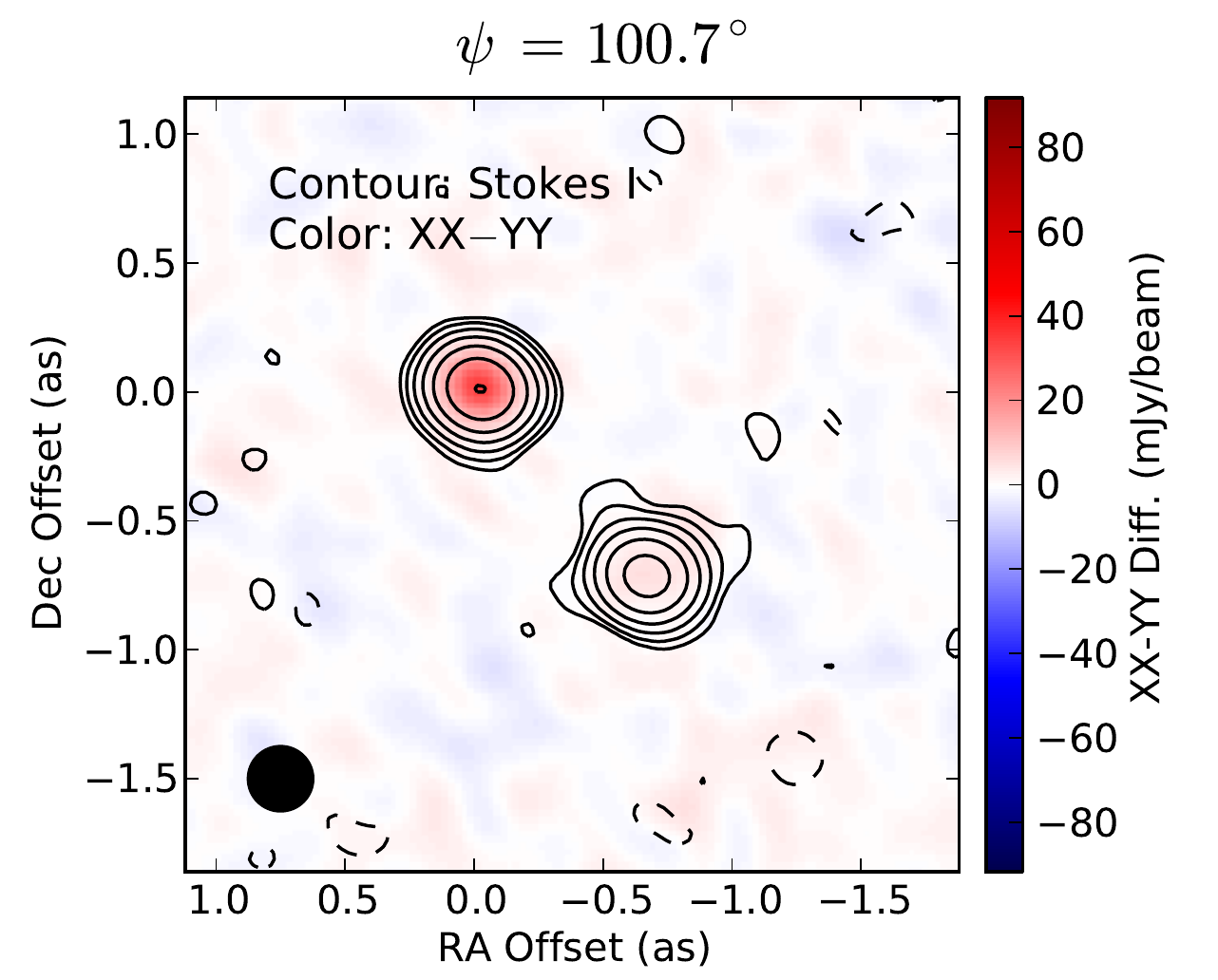}\hspace{-0.3cm} \vline
\includegraphics[width=6.4cm]{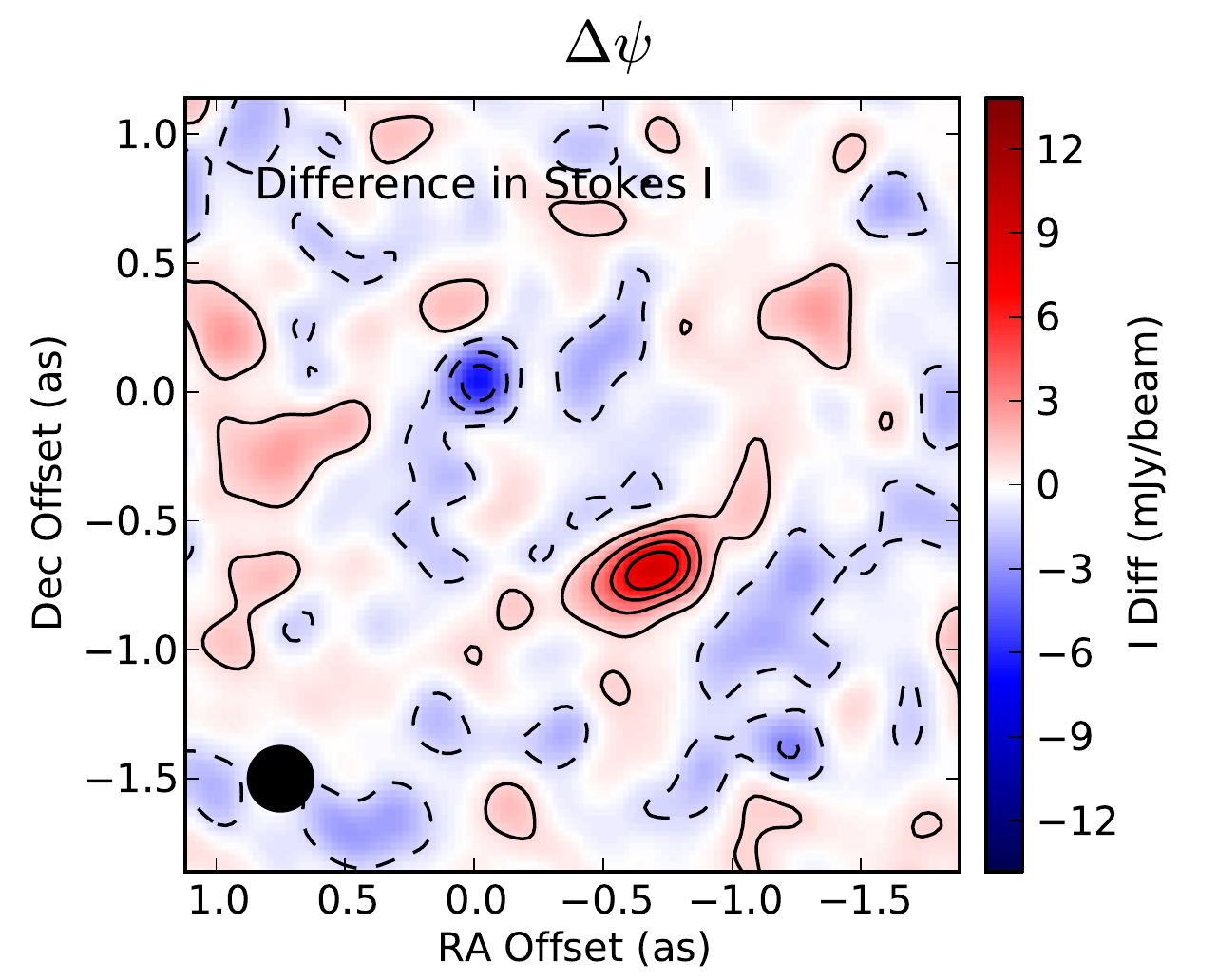}
\caption{Left and center: Band~9 Stokes I images (contours) and differences between images from polarizers XX and YY (colors) for two parallactic angles (at the start and end of the observations). The contours are spaced logarithmically between 0.5\% (dashed contour is at $-0.5$\%) and 95\% of the absolute Stokes I peak (i.e., the peak intensity among all observations). Right: difference in Stokes I between the first and last time bins (contours are linearly spaced at intervals of 0.2 times the maximum difference). The FWHM of the restoring beam is shown in the lower left side of each image (forced to 0.25\,as in all images).}
\label{MapsFig}
\end{figure*}

In Fig.\ref{MapsFig} we show Band~9 images for two values of the parallactic angle (i.e., the first and the last time bins in the observations). The polarization signal in the NE (as well as its dependence on $\psi$) can be clearly seen from the difference between the XX and YY images.

The flux-density ratios obtained for each polarizer separately, $R_{XX}$ and $R_{YY}$ (Fig.\ref{B9Fig}c), clearly encode the polarization signal from the NE, since the SW is nearly unpolarized. We note that the YY flux densities in the SW seem to be slightly lower than those in XX at all times. This may be related to a small gain-amplitude bias between X and Y at all antennas. However, this systematic does not affect the flux-density ratios, which are, by construction, free of any bias in the absolute flux-density calibration, as well as any instrumental and/or atmospheric effects \citep[e.g.,][]{IMV2013, IMV2016b}. For instance, the flux density clearly decreases for the NE and SW images around $\psi \sim 20$\,deg. (Fig.\ref{B9Fig}a,b), which does not appear in the flux ratios (Fig.\ref{B9Fig}c) and therefore indicates an instrumental/atmospheric origin. The NE/SW flux-density ratios in Stokes I show a clear change in all the observations (this is directly related to changes in the SW flux density). On the other hand, the ratios in XX and YY are symmetrically distributed around those in I, which occurs when at least one image in the lens is polarized \citep{IMV2015}. In Fig.\ref{MapsFig}  we also show the difference in Stokes I between the first and last time bins. The different sign in the time evolution of the NE and SW flux densities is clear evidence of the fast changes found in the NE/SW flux-density ratio.

Last but not least, we show the variations of the polarization ratio, $R_{pol}$, in Fig.\ref{B9Fig}d. The quantity $R_{pol}$ is independent of either gain-amplitude biases or variability in Stokes I at any (or both) of the images \citep{IMV2016}, so that it only encodes information about the difference in polarization between the two lensed images. In this case, since the SW image is unpolarized, $R_{pol}$ is directly related to be absolute polarization of the NE image.

In summary, the NE image shows a relatively stable flux density over the duration of the Band~9 observations, but with a clear polarization signal, while the SW image appears nearly unpolarized, but shows a steady increase in flux density of $\sim 10$\% in one hour.

\subsection{Band 7 and 8 data} \label{sec:resB78}

\begin{figure}
\centering
\includegraphics[width=9cm]{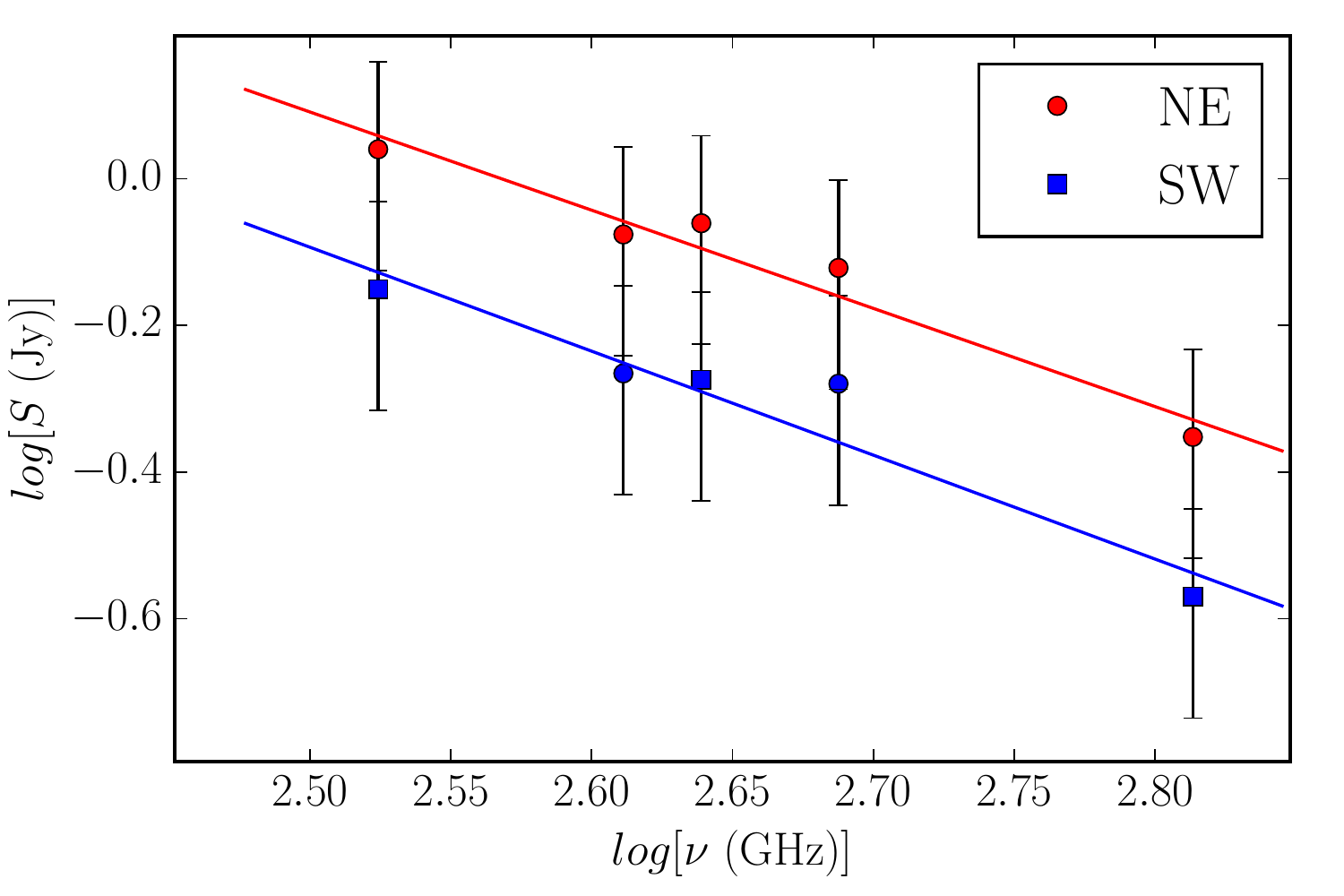}
\caption{Spectrum of the NE and SW image, taken from Bands 7 to 9.}
\label{SpecFig}
\end{figure}

In order to complement the Band~9 results discussed in the previous section, we have retrieved all other ALMA data observed in May-June~2015, and performed the same polarimetric analysis. We show the spectrum of the two lensed images in Fig.\ref{SpecFig}. Both spectra can be fit with the same spectral index, $\alpha = -1.2 \pm 0.2$, taking the flux density as $F_\nu = F_0 \left ( \frac{\nu}{\nu_0} \right ) ^{-\alpha}$. 

\begin{figure*}
\centering
\includegraphics[width=19cm]{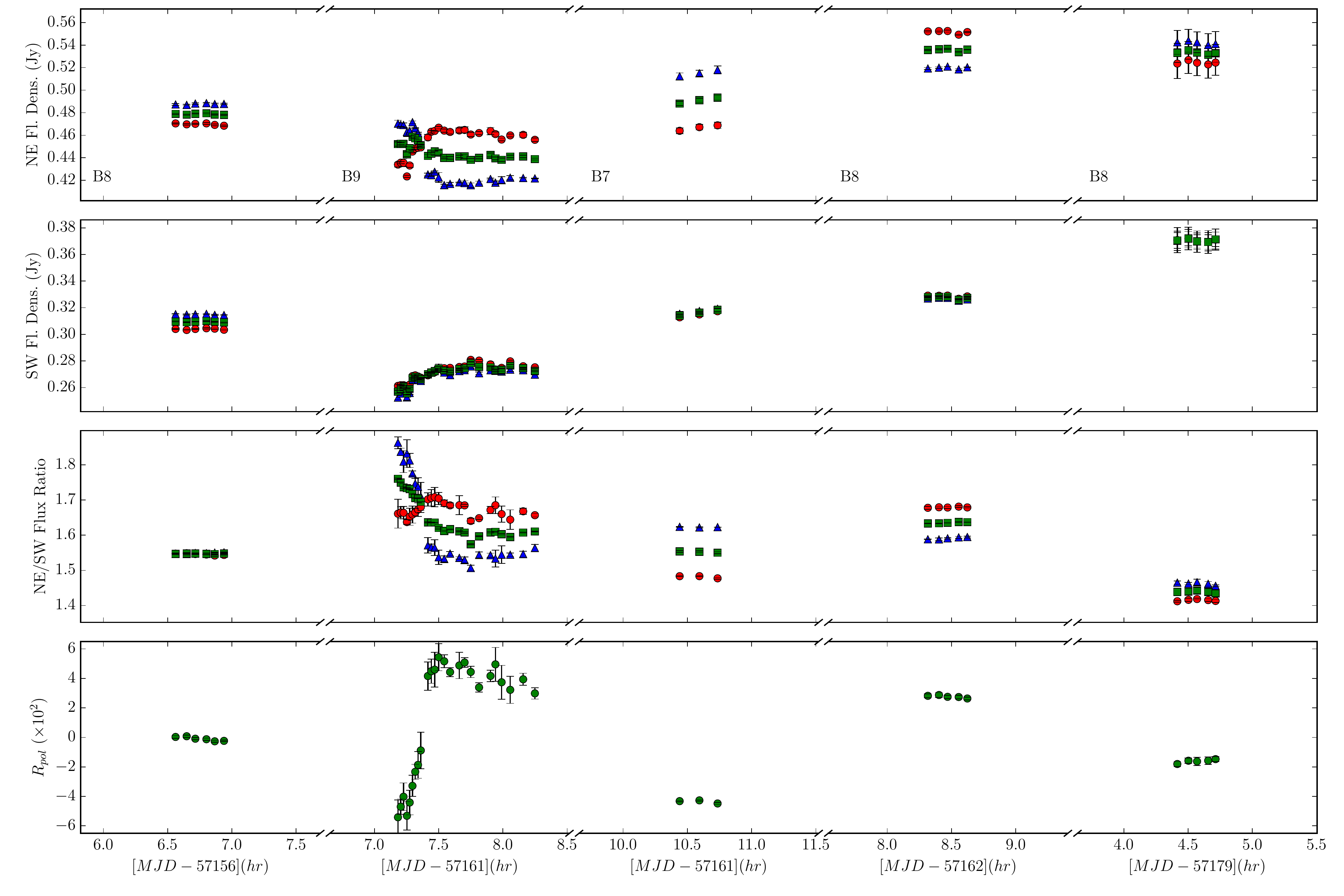}
\caption{Same quantities as in Fig.\ref{B9Fig} for all the ALMA data observed in May-June 2015, but as a function of time and with the flux densities normalized to a common frequency of 650\,GHz, assuming a fixed spectral index of $\alpha=-1.2$ (see Sect. \ref{sec:resB78}). The band corresponding to each epoch is shown in the top plots.}
\label{TimeFig}
\end{figure*}

The results of the polarimetry analysis are presented in Fig.\ref{TimeFig}. For clarity, we have scaled the flux densities of the NE and SW images to their values normalized to a common frequency of 650\,GHz (using the fitted $\alpha = -1.2$). Unfortunately, the available data are sparse in time and only cover a time range shorter than the time delay ($\Delta t \sim 27$~d) of the system. They also only cover a very narrow parallactic angle range each. Nevertheless, they shed some light on the intrinsic activity of the quasar.

The first epoch (MJD 57156; 2015 May 14) shows a flux ratio $R \sim 1.5$, close to the expected quiescent ratio (i.e., the lens geometric magnification ratio between the two images), as well as a negligible polarization ratio, $R_{pol} \sim 0$. The two images thus seem to be in the same polarization state (i.e., either both images are unpolarized or the fractional polarization is the same for both images, including the same EVPA). 

The next epoch (MJD 57161; 2015 May 19) corresponds to the Band~9 observations that we described in the previous section. The NE image is clearly polarized, judging from the $\psi$-dependent flux-density ratio between in XX and YY images. On the other hand, the SW image does not show any polarization signal: the flux density of SW in both polarizations (XX and YY) evolves in time in the same way.

In the following visit (2015 May 19 in Band~7, i.e., just about three hours after the Band~9 observations), the NE is still clearly polarized, whereas the SW image remains unpolarized. The fact that we obtain the same qualitative behavior from data taken with two independent receiver bands and within a few hours indicates that our technique is robust. The polarization ratio, $R_{pol}$, in Band~7 has the opposite sign of the polarization in Band~9, while the $\psi$ angle is very similar in the two observations. The reason for this difference in $R_{pol}$ is that the feed angle of the Band~7 receivers, in the frame of the antenna mounts, is different than that of the Band~9 receivers. If we take into account the difference between the feed angles of the receivers (see Table \ref{tab:summary-obs}), the Band~7 and Band~9 polarization ratios do agree with one single self-consistent source-polarization model (Fig.\ref{RMFitFig}, top).

On the following day (MJD 57162; 2015 May 20), observations were taken in Band~8 at a $\psi$ angle very similar to that at the end of the Band~9 observations. The difference in feed angles between these two bands is 180\,deg., which (in terms of differential polarimetry) does not introduce any rotation in $R_{pol}$. We recovered very similar flux-density ratios at both bands and all polarizations, as well as the same $R_{pol}$. This again confirms the robustness of our differential-polarimetry technique.

In the final epoch (MJD 57179; 2015 June 6), the Band~8 observations recover a very similar flux density for the NE image as in the previous visit, whereas the SW flux density has increased notably (note also the lower flux density ratio, around 1.45). Regarding the source polarization, $R_{pol}$ has a lower value, which may be related either to a lower fractional polarization in the NE image or to projection effects that are due to the different $\psi$ angle, and/or variations in the EVPA during the two-week time span between the observations.

\section{Discussion}

\subsection{Polarimetry and submm Faraday rotation}

At any given frequency, $\nu$, the polarization ratio, $R_{pol}$, is a function of the parallactic angle, $\psi$, the feed rotation at the antennas, $\beta_\nu$, and an angle $\phi_0$, related to the difference between the EVPAs of the lensed images \citep{IMV2015,IMV2016}:

\begin{equation}
R_{pol} = p_d \cos{\left[2(\psi + \beta + \phi_0)\right]},
\label{RpolSine}
\end{equation}

\noindent where $p_d$ is the differential polarization between NE and SW (or the absolute fractional polarization in NE, given that SW is unpolarized). In our case, since the polarization of SW is negligible, the differential polarization corresponds to the absolute polarization of NE \citep{IMV2016}. The best-fit model given by Eq.~\ref{RpolSine} is shown in Fig.\ref{B9Fig}, overplotted on the observed $R_{pol}$ values in Band~9. We estimate a fractional polarization of $p_d = (4.9 \pm 0.1)$\% and an EVPA of $(80.3 \pm 1.3)$\,deg. for the NE image.

\begin{figure}
\centering
\includegraphics[width=9cm]{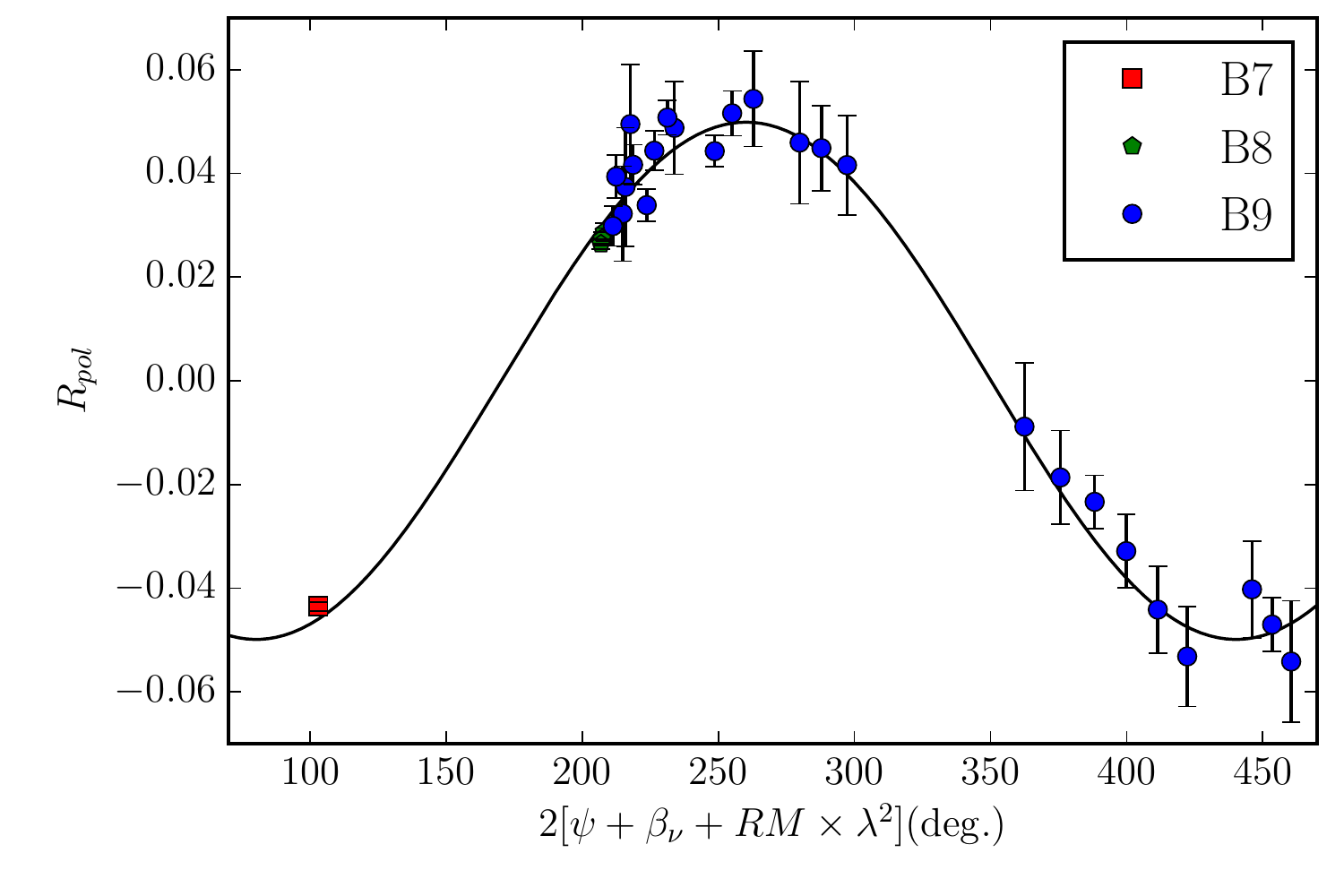}
\includegraphics[width=8.5cm]{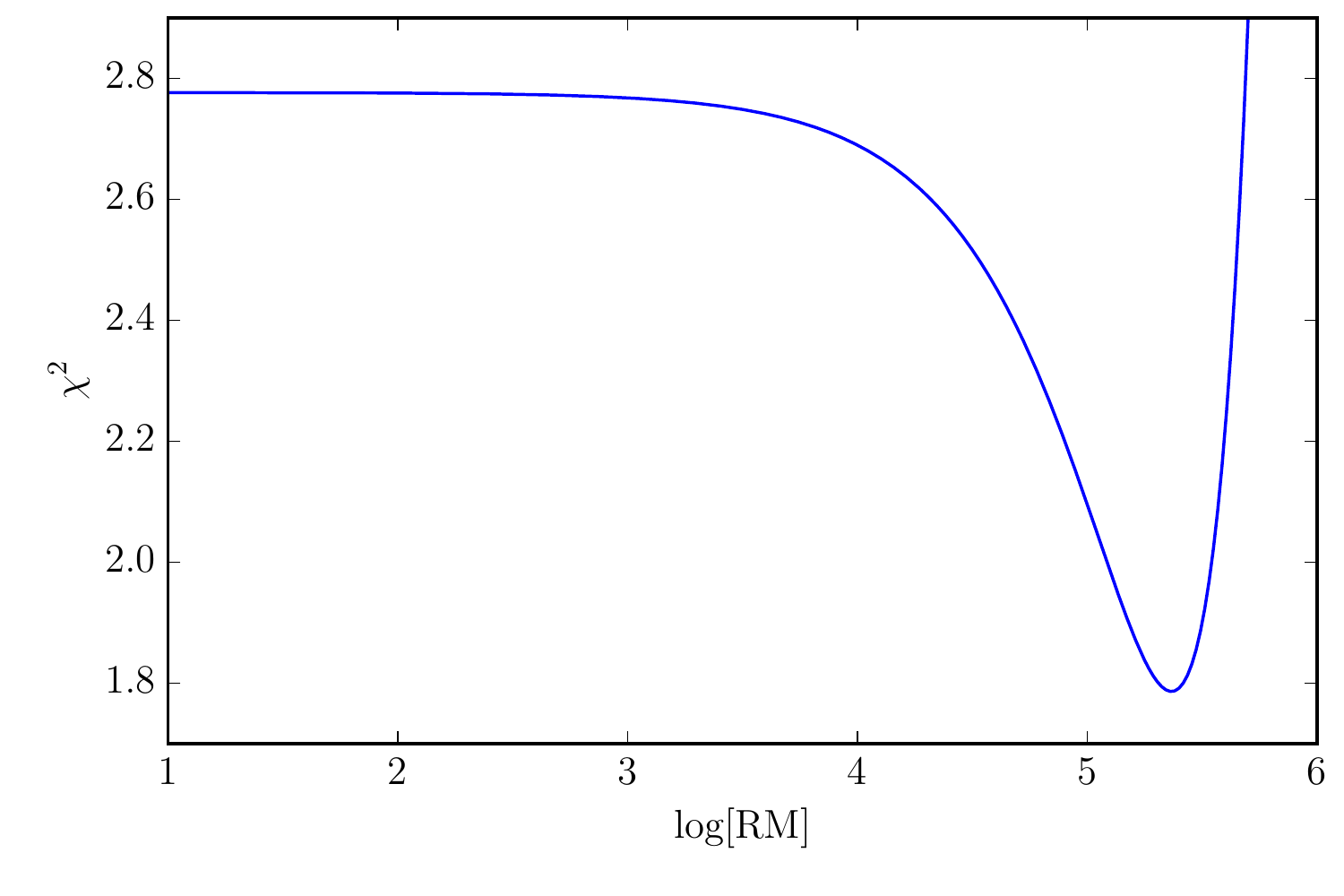}
\caption{Top: Data at Bands 7, 8, and 9, observed between May 19 and May 20, together with the best-fit rotation-measure model. Bottom: $\chi^2$ as a function of $RM$ (log scale in base 10).}
\label{RMFitFig}
\end{figure}

Given the good parallactic-angle coverage of the Band 9 observations and the quality of the fit of the $R_{pol}$ model (Fig.\ref{B9Fig}(d)), we can fix the differential polarimetry parameters in Band 9 (i.e., $p_d$ and $\phi_0$ in Eq.~\ref{RpolSine}) and fit for $RM$ in the NE image by adding the $R_{pol}$ measurements in Bands 7 and 8. The $R_{pol}$ model for the multi-frequency data is given by \citep{IMV2015}

\begin{equation}
R_{pol}^{\lambda} = p_d \cos{\left(2(\psi + \beta + \phi_0 + RM(\lambda^2-\lambda_0^2))\right)},
\label{RpolSineRM}
\end{equation}

\noindent where $\lambda_0$ is the central wavelength at Band 9 and $\lambda$ is the wavelength at which $R_{pol}^{\lambda}$ is measured. Combining the observations from May 19 and 20 (summarized in Table \ref{tab:summary-obs}) and assuming a negligible change in the EVPA during this time range ($\sim$ 1 day), we estimate $RM = (2.3 \pm 0.6)\times 10^5$\,rad\,m$^{-2}$. The data and best-fit model are shown in Fig.\ref{RMFitFig}. Short-timescale EVPA variability could affect our estimates of the rotation measure by $\delta RM = \delta {\rm EVPA}/\Delta (\lambda ^2)$, where $\Delta (\lambda)^2$ is the difference in wavelength squared between measurements. We note, however, that the observations at the extreme frequencies (Band~7 and 9), which provide the strongest constraint to the $RM$, were taken within only three hours. Our measured $RM$ could then be mimicked by an EVPA change of a few degrees within three hours (i.e., a rate of $\sim 20-30$~degrees per day), although this scenario would be at odds with the position of the Band~8 point (observed one day after) in Fig.\ref{RMFitFig}.

We note that the fitted $RM$ is far lower than the values reported by \cite{IMV2015}. The maximum EVPA rotation across Bands 7 to 9, corresponding to the fitted $RM$, is only  about $\sim 4$\,deg. The detection of $RM$ in these data is thus only tentative (the fitted value is still compatible with zero at a 4$\sigma$ level). The main reason for the high relative uncertainty in $RM$ is the poor parallactic-angle coverage of the data at Bands 7 and 8 as compared to Band 9, which does not allow us to use the $\psi$ dependence of $R_{pol}$ (i.e., the local slope of the sinusoid in Eq.~\ref{RpolSineRM}, at each frequency $\nu$) to derive a more accurate estimate of the EVPA differences among the frequencies.

\subsubsection{Spectral behavior of $R_{pol}$}

One solid conclusion from the previous subsection is that the $RM$ in May 2015 is much lower than the values on 2012/2014 reported by \cite{IMV2015}. If we were to have the same $RM$ in the observations reported here, the differences in EVPA from Band 7 to 9 would have been as large as 600\,deg. (i.e., almost two turns in phase). The maximum $RM$ reported by \cite{IMV2015} was so high that $R_{pol}$ changed even across the spectral windows within each tuning (see their Fig.1). This spectral dependence of $R_{pol}$ was indeed the main constraint on the order of magnitude of the $RM$ reported in \cite{IMV2015}; the effect of the feed angles among the bands was secondary and only affected the $RM$ estimates by a factor 2$-$3 \citep{IMV2016}. 

\begin{figure}
\centering
\includegraphics[width=9cm]{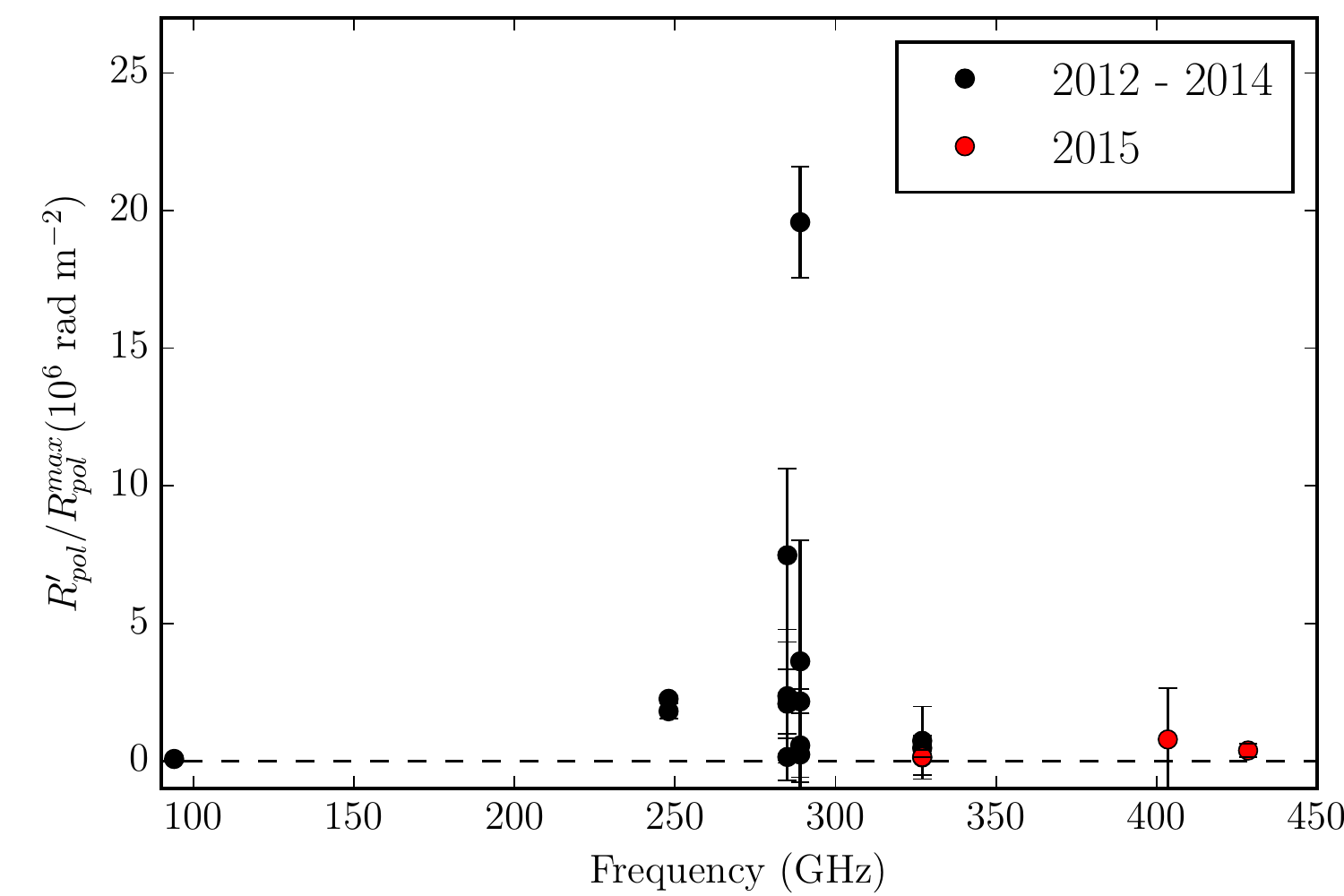}
\caption{$\frac{1}{R_{pol}}\frac{dR_{pol}}{d\lambda^2}$ as a function of frequency, fitted among the spectral windows of each frequency configuration.}
\label{SlopesFig}
\end{figure}

However, we did not find this $R_{pol}$ behavior in the 2015 data (see  our Fig.\ref{SlopesFig}). This is another clear indication of the (much) lower $RM$ for these epochs. According to Eq.~\ref{RpolSine}, the change of $R_{pol}$ as a function of $\lambda^2$ (for a fixed $\psi$) is 

\begin{equation}
\frac{dR_{pol}}{d\lambda^2} = R'_{pol} = 2 RM\,p_d\sin{\left(2 RM(\lambda^2-\lambda_0^2) + \alpha \right)},
\label{dRpolEq}
\end{equation}

\noindent where $\alpha = 2(\psi + \beta + \phi_0)$. If we approximate the maximum observed $R_{pol}$ (in absolute values, $R^{max}_{pol}$) to its upper bound (i.e., $R^{max}_{pol} \sim p_d$) and $\lambda \sim \lambda_0$, we have

\begin{equation}
\frac{R'_{pol}}{R^{max}_{pol}} \sim 2 RM \sin{\alpha}.
\label{dRpolRMEq}
\end{equation}

Although this equation is independent of $p_d$, it only gives us a rough (order-of-magnitude) estimate of $RM$. The peak of $R_{pol}$ is indeed related to the point where the derivative is null, whereas the derivative is maximum when $R_{pol}$ is minimum (in absolute value). On the other hand, if we approximate $R'_{pol}/R^{max}_{pol} \sim 2 RM$, we assume that the measurements are taken at $\alpha \sim \pi/4$. 

In other words, we cannot derive the exact value of $RM$ from a snapshot observation alone, even if we detect a variation of $R_{pol}$ with $\lambda^2$. To retrieve the $RM$, we also need a good coverage of parallactic angle, $\psi$. We can, however, still constrain the order of magnitude of $RM$ from Eq.~\ref{dRpolRMEq}.

In Fig.\ref{SlopesFig} we show the slopes of the relative change in $R_{pol}$ (i.e.,  $R'_{pol}/R^{max}_{pol}$), for the epochs reported by \cite{IMV2015} and for the data reported here. These values have been obtained by fitting a linear model in $\lambda^2$ space to each frequency configuration at each epoch. The difference between epochs is clear: the slopes in 2012$-$2014 are higher than the values obtained in 2015. For the epochs reported in \cite{IMV2015}, there are some detections of non-zero slopes at several $\sigma$, whereas all measurements for 2015 are compatible with zero. 

We also note that the lowest frequencies reported here correspond to Band~7 ($\sim$330\,GHz), whereas \cite{IMV2015} also reported observations in Band~6 ($\sim$250\,GHz), for which the fractional bandwidth is $\sim$30\% wider (and the $RM$ accordingly imprints larger $R_{pol}$ variations across the frequency coverage).

High $RM$ variability \citep[e.g.,]{Asada08,Lico14,Beuchert2018}, on the order of a year in blazars like Mrk\,421, and even $RM$ with sign reversals \cite[e.g.,][]{Sullivan09,Gomez11,Gomez17} have been observed in several AGN. The dramatic changes that we have observed in the $RM$ of \PKS1830, covering several years, are therefore not surprising.

\subsection{Short-timescale variability}

The NE/SW flux-density ratio in Stokes I, $R_I$, decreases in time during the Band~9 observations, from a value of $\sim1.75$ (at 7:10\,UT) to $\sim1.60$ (at 8:17\,UT). This is a change of $\sim10$\% in $R_I$, almost completely related to the change in flux density of SW (Figs.\ref{B9Fig}b and \ref{TimeFig}), although there is signature of weak variability in NE as well (Figs.\ref{B9Fig}a and \ref{MapsFig} right). Unfortunately, there is no way to decouple more precisely how much of the $R_I$ variation is due to each one of the two lensed images. 

If the NE flux density did not vary, so that the residual in the difference in Stokes I between the first and last time bins (blue spot seen in Fig.\ref{MapsFig} right) is only due to instrumental/atmospheric instability during the observations, then the SW image varied its flux density at an average rate of $\sim 15$~mJy per hour (i.e., $\sim10$\%, the rate for $R_I$). This value should be taken as an upper limit to the actual intrinsic variability of SW, because we cannot rule out any simultaneous anticorrelated variation of NE either. 

In contrast, the Band~7 and 8 datasets do not show obvious short-term variability (Fig.\ref{TimeFig}). However, their observing time span was typically a factor 2$-$3 shorter than in the Band~9 data, which makes it more difficult to detect significant changes.

Intra-hour variability in the mm/submm emission from \PKS1830 was previously reported by \cite{IMV2013} for Band~7 (300\,GHz), based on rapid (intra-hour) variations of about 3\% in the NE/SW intensity ratio. According to the model proposed there, the Band~7 observations were taken very close in time to an event of matter injection into the jet base, seen in the SW image (the same event, but occurring $\sim 27$~days before in the NE image, was responsible for the varying ratios at the previous epochs). Unfortunately, the time and frequency coverage of the observations reported here is not good as those presented by \cite{IMV2013}, which prevents us from conducting a deeper study of the intrinsic variability of the source. Nevertheless, this new case of episodic variability seen in Band~9 establishes the recurring submm activity of \PKS1830, which makes it the ideal target for a study of AGN variability {at unprecedented accuracy in the submm.

According to the Fermi-LAT light-curve of \PKS1830, there was no apparent gamma-ray activity at the time of our ALMA observations. We note, however, that the gamma-ray emission started to steadily increase shortly after our observations, covering a factor $\sim5$ in photon counts in about 200 days (taken from the Fermi monitoring database\footnote{\texttt{https://fermi.gsfc.nasa.gov/ssc/data/access/lat/msl\_lc}}).

\section{Summary} \label{sec:summ}

We reported the analysis of ALMA continuum data of the quasar \PKS1830\ at submillimeter wavelengths (Bands 7, 8, and 9, with a sky frequency range from 330 to 650\,GHz, i.e., up to 2.3~THz redshift-corrected frequencies), taken in May-June 2015. The Band~9 data, which cover a wide parallactic-angle range, show a clear polarization signal, which motivated us to conduct this work. Applying a differential-polarization method \citep{IMV2016}, we were able to retrieve some polarization information from the two lensed images of the quasar, even though the data were not taken in full-polarization mode. Our results can be summarized as follows:

\begin{itemize}

\item We find a fractional polarization of $\sim 5$\% for the NE image of \PKS1830, while the SW image appears nearly unpolarized. Since both images have a time delay of about 27~days, this is evidence of polarization variability on a timescale of a few weeks.

\item Comparing the Band~9 observations with concomitant Band 7 and 8 data, we find self-consistent polarimetry results for the two lensed images. This self-consistency allows us to constrain the Faraday rotation measure, $RM$, up to a few times $10^5$\,rad\,m$^{-2}$, at submm wavelengths. However, the RM accuracy is limited, which prevented us from performing a detailed analysis of its physical implications. In any case, this value is much lower than the $RM$ previously reported for this source in 2012 to 2014 \citep{IMV2015}, and it indicates strong variability in the source polarization properties.

\item In addition, we find evidence of rapid flux-density variability ($\lesssim 10$\%) in the SW image within the approximately one-hour time span of the Band~9 observations. Such intra-hour variability is absent at other bands (the data were taken at different times and with a shorter observing time span, however), but has been previously noted in 300~GHz data \citep{IMV2013}.

\item Altogether, \PKS1830 shows evidence for submm variability on timescales of hours to years.

\end{itemize}

The current generation of mm/submm instruments ushers in a golden age for high-precision submm-wave AGN polarimetry. The use of the differential method, which is virtually free of calibration issues, pushes the accuracy of submm polarimetry even further, providing us with the opportunity of probing the magneto-ionic conditions in AGN regions extremely close to their central engines. In this context, \PKS1830\ has already proven to be the ideal target for this type of studies.

\begin{acknowledgement}
We thank the referee for constructive comments.
This paper makes use of the following ALMA data: \\
ADS/JAO.ALMA\#2013.1.00020.S,\\
 ADS/JAO.ALMA\#2013.1.00296.S,\\
and ADS/JAO.ALMA\#2013.1.01099.S.\\
ALMA is a partnership of ESO (representing its member states), NSF (USA) and NINS (Japan), together with NRC (Canada) and NSC and ASIAA (Taiwan) and KASI (Republic of Korea), in cooperation with the Republic of Chile. The Joint ALMA Observatory is operated by ESO, AUI/NRAO and NAOJ. This research has made use of NASA's Astrophysics Data System.
\end{acknowledgement}

\end{document}